\documentclass[preprint,apl,showpacs,showkeys,preprintnumbers,amsmath,amssymb]{revtex4}
\usepackage{graphicx}
\usepackage{dcolumn}
\usepackage{bm}

\begin{document}
\preprint{Wang et. al., submitted to APL}
\title{Ti-rich and Cu-poor grain-boundary layers of CaCu$_3$Ti$_4$O$_{12}$ detected by x-ray photoelectron spectroscopy}
\author{C. Wang, H. J. Zhang, P. M. He, and G. H. Cao\footnote[1]{To whom correspondence should be addressed (ghcao@zju.edu.cn)}}
\affiliation{Department of Physics, Zhejiang University, Hangzhou
310027, People's Republic of China}

\date{\today}

\begin{abstract}
Cleaved and polished surfaces of CaCu$_3$Ti$_4$O$_{12}$ ceramics have been investigated by x-ray photoelectron
spectroscopy (XPS) and energy dispersive x-ray spectroscopy (EDX), respectively. While EDX technique shows the
identical CaCu$_3$Ti$_4$O$_{12}$ stoichiometry for the two surfaces, XPS indicates that the cleaved surface with
grain-boundary layers is remarkably Ti-rich and Cu-poor. The core-level spectrum of Cu 2$p$ unambiguously shows
the existence of monovalent copper only for the cleaved surface. Possible grain-boundary structure and its
formation are discussed.
\end{abstract}

\pacs{79.60.-i, 61.72.Mm, 77.84.Dy}

\keywords{x-ray photoelectron spectroscopy, grain boundaries, CaCu$_3$Ti$_4$O$_{12}$}

\maketitle

CaCu$_3$Ti$_4$O$_{12}$ (CCTO) has recently attracted considerable interest due to its extraordinarily high
dielectric permittivity ($\sim10,000$) at low frequencies over a wide temperature range (100 K $\sim$ 400
K).\cite{Subramanian,Homes} The giant dielectric phenomenon has been primarily elucidated as an extrinsic effect
in terms of an internal-barrier-layer-capacitor mechanism.~\cite{Sinclair} It was suggested~\cite{Sinclair} and
then confirmed~\cite{Chung} that grain boundary (GB) was the internal barrier for CCTO ceramics. As for CCTO
crystals which show even larger dielectric permittivity,~\cite{Homes} internal barriers were still inferred from
impedance spectroscopy measurement.~\cite{Li} In fact, CCTO ceramics contains domain boundaries as well as grain
boundaries, revealed by scanning electron microscopy and high-resolution transmission electron
microscopy.~\cite{Fang} As a result, the detailed dielectric responses of CCTO ceramics can be well interpreted
by using a double-barrier-layer-capacitor model.~\cite{Cao}

So far, however, the structure of the internal barriers (grain boundaries and domain boundaries) remains
unclear. It was initially suggested that twin boundary was the possible barrier layer.~\cite{Subramanian} A
structural model of planar defects due to a twining parallel to $\{100\}$ planes was recently
proposed,~\cite{Whangbo} yet it needs further experimental support. According to a detailed TEM
investigation,~\cite{Wu} such twin domains could not be detected in single crystals or polycrystallines,
instead, high density of dislocations and cation-disorder-induced planar defects were observed. Very recently,
x-ray diffraction under extremely high hydrostatic and uniaxial compression suggested that CCTO ceramics was
composed of grains with stiffer shells and softer cores.~\cite{Ma} If so, the GB layers should be different from
the grain interiors in structure and composition. However, another recent report\cite{MLi} on Mn-doping effect
proposed that the grain and GB regions in CCTO ceramics might consist of the same phase but with slightly
different compositions.

In this Letter, x-ray photoelectron spectroscopy (XPS) and energy dispersive x-ray spectroscopy (EDX) were
employed to detect the possible differences between a polished surface (PS) and a cleaved surface (CS) of CCTO
ceramic samples. For a CS, GB layers remain on the surface because of the relatively weak linkage between
grains. In the case of a PS, on the other hand, grain interiors are exposed on the surface. Because the
detecting depth of XPS is $\sim$ 1 nm in most cases,~\cite{Tanuma} XPS actually reveals the information of
ultra-thin surface layers. In comparison, the information depth of EDX is commonly at the micron scale, thus EDX
measures the bulk composition. By examining the CS and PS layers with the two techniques, one may obtain the
compositional and structural information, especially for the GB layers.

The CCTO ceramic samples were prepared by conventional solid-state reaction using the powdered chemicals of
TiO$_2$ (99.99\%), CaCO$_3$ (99.99\%), and CuO (99.99\%). The starting materials were weighed according to the
stoichiometric ratio and mixed thoroughly in an agate mortar. The mixed powder was calcined at 1273 K for 12 h
in air. This procedure was repeated for three times to ensure that the samples were in single CCTO phase. Then
the calcined powder was pressed into a disk ($\phi$12 mm$\times$2 mm) and a rod ($\phi$12 mm$\times$15 mm)
respectively. The pressed specimens were finally sintered in air at 1353 K for 24 h followed by furnace-cooling
to room temperature. X-ray diffraction identified single phase for the two specimens. Dielectric measurement
with an Agilent 4284A precision LCR meter confirmed that the samples showed the property of giant dielectric
permittivity as reported elsewhere.~\cite{Subramanian,Cao}

Prior to the XPS and EDX measurements, the as-prepared CCTO samples were treated to make a PS and a CS,
respectively. For the PS sample, the sintered disk was polished by using CeO$_2$ fine powder, followed by
removing the remaining CeO$_2$ with a mixture of nitric acid and hydrogen peroxide. Then it was in turn cleaned
in distilled water, ethanol, and acetone with a ultrasonic cleaner. The CS sample was obtained simply by
cleaving the rod. As soon as the surface was made, the specimen (mounted on a sample holder) was transferred
into the ultra-high vacuum system equipped with an x-ray generator (Mg K$\alpha$, 1253.6 eV) and an Omicron
EAC2000-125 analyzer. The shift of core-level spectra due to the charging effect was calibrated using the
contaminated C $1s$ peak located at 284.6 eV.~\cite{Dobler} The XPS intensity was calculated based on the areas
of the related peaks. After the XPS experiment, the identical samples were also examined by employing a
field-emission scanning electron microscope (SEM, SIRION FEI, Netherlands) equipped with a Phoenix (EDAX) x-ray
spectrometer. The samples were coated with very thin layer of gold before they were placed into the SEM chamber.

\begin{figure}
\includegraphics[width=7.5cm]{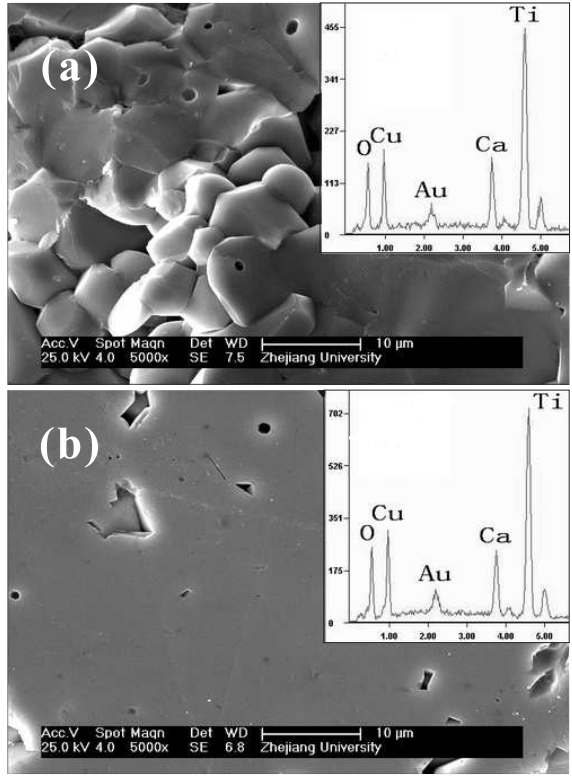}
\caption{SEM images of cleaved surface (a) and polished surface (b) of CaCu$_3$Ti$_4$O$_{12}$ ceramics. The
insets show the corresponding EDX spectra, respectively. Both of the surfaces were gold-coated prior to the SEM
observation.}
\end{figure}

Fig. 1 shows the SEM images of the cleaved and polished surfaces of CCTO ceramics. The CS image shows
grain-packed morphology with the grain size of $\sim$ 6 $\mu$m. One can see that the cleaving takes place mostly
at the grain boundaries. For the PS image, a flat surface is shown except for some cavities. This indicates that
grain interiors show up due to the polishing. It is also noted that both surfaces have very similar EDX spectra,
as shown in the insets of Fig. 1. Quantitative analysis indicates that the atomic ratios (Ca:Cu:Ti) for CS and
PS are 1.0:3.1:4.0 and 1.0:3.0:3.8, respectively, consistent with the stoichiometric ratio of CCTO within the
experimental errors ($\sim$ 5\%).

Fig. 2 shows the core-level spectra of the cleaved and polished surfaces of CCTO ceramics. Although there is no
obvious shift for the positions of the core-level peaks, the relative intensities vary remarkably. As can be
seen in the panel (a), the intensities of Ca $2p$ peaks are similar for the two surfaces. In comparison, the
intensity of Cu Auger peak is remarkably weaker for the CS. Accordingly, the intensity of Cu $2p$ peaks in the
panel (b) is also proportionally weaker for the CS. For the Ti $2p$ peaks in the panel (c), however, the
intensity is substantially higher for the CS. In the case of O $1s$ peak, it is difficult to make a comparison
because of the disturbance of surface contamination, especially for the PS sample.

\begin{figure}
\includegraphics[width=7.5cm]{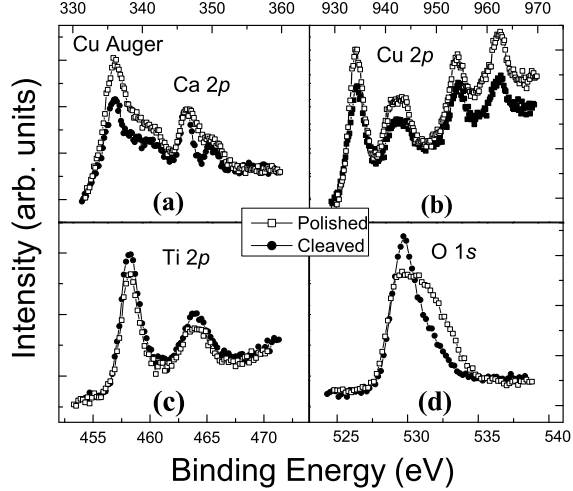}
\caption{Core-level spectra of Ca $2p$ (a), Cu $2p$ (b), Ti $2p$ (c) and O $1s$ (d) for the cleaved and polished
surfaces of CaCu$_3$Ti$_4$O$_{12}$ ceramics.}
\end{figure}

Here we make a quantitative comparing analysis so that the element-specific and instrumental parameters do not
need to be concerned. The cation ratio of the PS is assumed to be $N^{ps}_{Ca}:N^{ps}_{Cu}:N^{ps}_{Ti}=1:3:4$,
because the grain interiors were exposed adequately. Consequently, with the XPS data of the PS as a reference,
the composition of the CS was easily determined as $N^{cs}_{Ca}:N^{cs}_{Cu}:N^{cs}_{Ti}=1.0:2.3:4.7$ (the
measurement error, mainly coming from the determination of the peak area, is no more than 5\%). Since the
information depth is only about 1 nm, the XPS result of the CS reveals the information of the GB layers.
Therefore, we conclude that the GB layers are Ti-rich and Cu-poor, compared with the CCTO stoichiometry.

Fig. 3 separately shows the Cu $2p$ core-level spectra of CCTO ceramics. As can be seen, there is a shoulder at
lower energy side of the main peak of $2p_{3/2}$ only for the CS specimen. By separating the peaks one obtained
a small peak at 932.2 eV, suggesting the existence of Cu(I) at GB (similar peak separating for the PS specimen
was unsuccessful). Another evidence for the existence of Cu(I) comes from the intensities of the Cu $2p$
shake-up peaks. The CS sample shows relatively weak shake-up peaks, because Cu(I) has no such component.

\begin{figure}
\includegraphics[width=7.5cm]{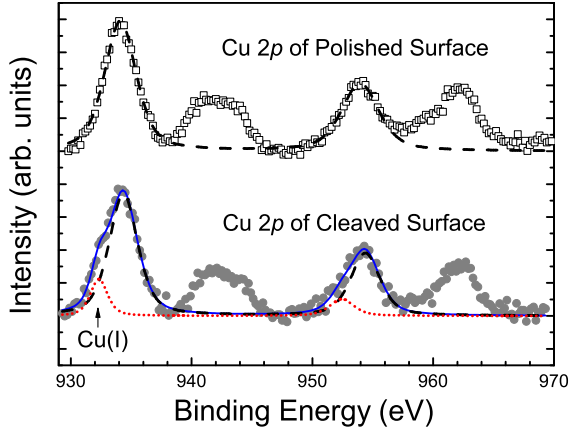}
\caption{Analysis on Cu $2p$ spectra of the cleaved and polished surfaces of CaCu$_3$Ti$_4$O$_{12}$ ceramics.
Note that the background was subtracted by using Shirley-plus-linear method.}
\end{figure}

One notes that the amount of excessive Ti is almost equal to the amount of the missing Cu for the CS. This
result suggests that some Ti may replace Cu in the GB layers. In fact, evidence of Ti on the Cu site was given
in nonstoichiometric Sr$_{0.946}$(Cu$_{2.946}$Ti$_{0.054}$)Ti$_4$O$_{12}$~\cite{Li2} and
Na(Cu$_{2.5}$Ti$_{0.5}$)Ti$_4$O$_{12}$.~\cite{Avdeev} In CCTO, however, the Ti on the Cu site is far too small
to be detected by refining site occupancies from neutron diffraction data.~\cite{Li2} Nevertheless, there is a
possibility for the GB in which significant amount of Ti occupies the Cu site. The Ti-for-Cu substitution
results in monovalent copper due to the charge neutrality, in agreement with the existence of Cu(I) only for the
GB layers.

With the clue of Ti on Cu site, Li et al.\cite{Li2} proposed an convincing explanation of how CCTO develops
conducting regions. At high temperature, Cu(II) reduces to Cu(I) accompanying with a charge compensation via a
slight substitution of Ti(IV) on Cu site, forming
Ca(Cu$^{2+}_{1-3x}$Cu$^{+}_{2x}$Ti$^{4+}_{x}$)$_3$Ti$^{4+}_4$O$_{12}$. Upon cooling, the Cu(I) converts to
Cu(II), liberating electrons into the Ti 3$d$ conduction band. Although this mechanism explains the conductivity
of CCTO grains, one could not understand the formation of internal barriers. We notice that the above mechanism
ignores the cation migrations during the cooling process. It is possible that the Ti at Cu site migrates to GB
layers and domain-boundary layers when cooling down, which forms a Ti-rich and Cu-poor barrier layer.

It is also noted that the structure model proposed by Wu et al.~\cite{Wu} is consistent with the result of
Ti-for-Cu substitution in GB layers. The planar defect model involves a lattice shift
$R=\frac{1}{4}[110]$.\cite{Note} Such a lattice shift naturally results in cation disorder in Ca/Cu site,
accommodating the Ti-for-Cu substitution. The planar defect may give rise to remarkable strain at the boundary,
which explains the thermal etching effect within the grains.~\cite{Fang} Furthermore, such a domain-boundary may
serves as a stiffer layer as suggested by the high-pressure x-ray diffraction result~\cite{Ma}. Finally, due to
the lattice discontinuity, ion-disorder and/or ion displacement, the interface may become a barrier layer
against the electron conduction.

In summary, we have revealed the subtle differences in composition of the bulk and grain boundary regions in
CCTO ceramics by comparing the core-level spectra of cleaved and polished surfaces. The grain boundary contains
remarkably more Ti and less Cu than the grain interior does. Moreover, only the grain boundary layer contains
monovalent copper. These results provide crucial insight for the origin of the special giant dielectric
phenomenon as well as the grain-boundary structure in CCTO system.

This work was supported by National Science Foundation of China (Grant No. 10274070).


\begin{thebibliography}{0}
\bibitem{Note}The original model contains an extra oxygen atom at the interface, what is more, two oxygen atoms are too much close. A more reasonable structure for the planar defect can be simply viewed as
$\frac{1}{4}[110]$ shear planes, involving edge-sharing TiO$_6$ octahedra.
\end{thebibliography}
\end{document}